\documentclass[prl,twocolumn,showpacs,amsmath,amssymb,superscriptaddress]{revtex4}
\usepackage{pifont,amssymb,epsf,graphicx}
\begin{document}
\title{Coulomb Interaction-induced Checkerboard Patterns in Disordered Cuprates}
\author{Degang Zhang}
\affiliation{Texas Center for Superconductivity and Department 
of Physics, University of Houston, Houston, TX 77204, USA}

\begin{abstract}

We study the effect of the Coulomb interaction on the local density of states (LDOS) 
and its Fourier component in disordered cuprates. It is shown that the Coulomb interaction  suppresses strongly the maximum value of the LDOS induced by the dopant impurity at each energy and expands significantly the Friedel oscillation in real space. The existence of the Coulomb interaction with a moderate strength yields an energy-dependent checkerboard LDOS modulation around the impurity, which is very different from that produced by pure quasiparticle interference. The orientation and transformation of the checkerboard pattern with energy and the relations among the modulation vectors, dopings and the bias voltages agree qualitatively with the recent STM experiments.  
 
\end{abstract}
\pacs{74.62.Dh, 74.72.-h, 74.25.-q}
\maketitle

It is well-known that at half filling cuprates are the Mott insulators, where the on-site Coulomb
interaction is larger than the kinetic energy of electron. When oxygens (holes) are doped into cuprates, 
the electrons can move in the CuO$_2$ plane and cuprates become strange metals. In other words, 
the "effective" on-site Coulomb interaction decreases.
With further increasing the concentration of holes up to about $5\%$, cuprates become superconducting 
and have a d-wave symmetry gap. So cuprate superconductors inevitably possess disorder due to the oxygen doping.
It is out of question that the electronic states in the CuO$_2$ plane are mainly determined by disorder and the Coulomb interaction.

Recently a series of the STM experiments have revealed the electronic states in both real and momentum spaces for the superconductor Bi$_2$Sr$_2$CaCu$_2$O$_{8+\delta}$ [1,2,3]. It was found that in real space the LDOS has a weak 
energy-dependent checkerboard pattern. At low bias voltages, the charge modulation with a period $\frac{2\pi}{q_B}$ orients along 45$^0$ to the Cu-O bonds. At high bias voltages well below the superconducting gap, the charge modulation with a period $\frac{2\pi}{q_A}$ is parallel to the Cu-O bonds. We note that $q_A$ are the wave vectors connecting the end points of the same Banana-like equal-energy contours while $q_B$ are the wave vectors connecting the nearest end points of the neighbor Banana-like equal-energy contours [1,2,4,5]. These two kinds of charge modulations coexist at intermediate voltages. With increasing the bias voltage or doping, $q_A$ becomes shorter while $q_B$ becomes longer [1,2]. In Ref. [3], McElroy {\it et al.} studied in detail the the nanoscale electronic disorder induced by the oxygen dopant atoms. It was indeed verified that these dopant defects generate the LDOS modulations and also suppress superconducting coherence peaks in Bi$_2$Sr$_2$CaCu$_2$O$_{8+\delta}$. 

A lot of theoretical investigations have devoted to these fascinatingly experimental observations [4-13]. In momentum space, the relation among the modulation wave vectors, dopings and the bias voltages seems to be well explained by quasiparticle interference due to weak disorder. However, in real space, the LDOS near the impurity has a stripe-like structure along both $45^0$ to the Cu-O bonds and the Cu-O bonds  rather than the checkerboard pattern at each energy [9]. In addition, the Friedel oscillation only exists in a short distance ($\sim$10 lattice constant) away from the impurity site. For many random impurities, a checkerboard pattern and its transformation with energy and doping are observed theoretically. But the checkerboard pattern only covers a small indefective area and is different from that seen by the STM experiments, which expands into the bulk. Therefore, it is difficult to interpret simultaneously the experimental phenomena in both real and momentum spaces by using pure quasiparticle interference induced by disorder. 
What factor makes the LDOS at different energy have the two-dimensional period structure in the bulk? It is natural to consider the Coulomb interaction in the CuO$_2$ plane as a candidate because the undoped cuprates are the Mott insulators. In this paper, we investigate the influence of the Coulomb interaction on the Friedel oscillation in disordered cuprate superconductors. We shall see that the experimental phenomena are the results of disorder correlation mediated by the Coulomb interaction.

The Hamiltonian describing the scattering of quasiparticles from a single
defect with local modifications of both hopping and pairing parameters and the Coulomb interaction in a
d-wave superconductor can be written as
$$H=\sum_{{\bf k}\sigma }({\epsilon}_{\bf k}-
\mu){c}_{{\bf k}\sigma }^{+}{c}_{{\bf k}\sigma
}+\sum_{{\bf k}}{\Delta }_{{\bf k}}(c_{{\bf k}
\uparrow }^{+}c_{-{\bf k}\downarrow }^{+}+c_{-{\bf k}
\downarrow }c_{{\bf k}\uparrow })$$
$$+\sum_{<i,j>,\sigma }{\delta t}_{ij}c
_{i\sigma }^{+}c_{j\sigma }+\sum_{<i,j>}{\delta \Delta }
_{ij}(c_{i\uparrow }^{+}c_{j\downarrow }^{+}+c
_{j\downarrow }c_{i\uparrow })$$
$$ +(V_{s}+V_{m})c_{0\uparrow }^{+}c
_{0\uparrow }+(V_{s}-V_{m})c_{0\downarrow }^{+}
c_{0\downarrow}$$
$$+U\sum_{i}c_{i\uparrow }^{+}c_{i\uparrow }^{}c_{i\downarrow }^{+}c_{i\downarrow }
+\frac{1}{2}V\sum_{<i,j>,\sigma,\sigma^\prime}
c_{i\sigma }^{+}c_{i\sigma }^{}c_{j\sigma^\prime }^{+}c_{j\sigma^\prime },\eqno{(1)}$$
where $<i,j>$ denotes that the sites $i$ and $j$ are the nearest neighbor sites, $\mu $ is the chemical potential to be determined by doping, 
$\epsilon_{{\bf k}}=t_{1}(\cos {k}_{x}+\cos {k}_{y})/2+t_{2}\cos 
{k}_{x}\cos {k}_{y}+t_{3}(\cos 2{k}_{x}+\cos 2
{k}_{y})/2+t_{4}(\cos 2{k}_{x}\cos {k}_{y}+\cos 
{k}_{x}\cos 2{k}_{y})/2+t_{5}\cos 2{k}_{x} \cos 2
{k}_{y}$ with $t_{1-5}$=-0.5951, 0.1636, -0.0519, -0.1117, 0.0510
(eV). The band parameters are taken from those of Norman {\it et al}.
[14] for Bi$_{2}$Sr$_{2}$CaCu$_{2}$O$_{8+\delta }$, and the lattice constant 
$a$ is set as $a$=1. The order parameter away from the impurity is given by $
\Delta _{{\bf k}}=\Delta _{0}(\cos {k}_{x}-\cos {k}_{y})/2 $. $\delta t_{ij}=\delta t$ for $i=0$ or $j=0$ and 0 for others. $\delta \Delta_{ij}=\delta \Delta_1$ for $i=0$ or $j=0$, $\delta \Delta_2$ for $i=1$ or $j=1$, and $0$ for others. $U_s$ and $V_m$ are the nonmagnetic part and the magnetic part of an on-site potental, respectively. $U$ is the strength of the on-site Coulomb interaction while $V$ is that between the nearest neighbor sites.
 
The Hamiltonian (1) cannot be diagonalized exactly due to the presence of the Coulomb interaction. Because the charge modulations observed by the STM experiments are weak, here we are only interested in the case of weak to moderate $U$ and small $V$, where the charge density wave (CDW) and spin density wave (SDW) can be neglected [15]. In fact, the presence of CDW and SDW only produces an energy-independent peak at the modulation wave vectors in the Fourier component of the LDOS and doesnot affect the results in this paper [16-22]. 

Following the standard Green's function method [23], we can solve the Hamiltonian (1) in the absence of CDW and SDW. Then the LDOS measured by the STM experiments reads
$$\rho({\bf r},\omega)=-\frac{1}{N\pi}{\rm Im} \sum_{{\bf k},{\bf k}^\prime,\nu,\nu^\prime}\cos [({\bf k}-{\bf k}^\prime)
\cdot {\bf r}][(-1)^{\nu+\nu^\prime}\xi_{{\bf k}\nu}\xi_{{\bf k}^\prime\nu^\prime}$$
$$\times G^{\nu\nu^\prime}_{{\bf k}{\bf k}^\prime}(i\omega_n)-\xi_{{\bf k}\nu+1}\xi_{{\bf k}^\prime\nu^\prime+1}G^{\nu\nu^\prime}_{{\bf k}{\bf k}^\prime}(-i\omega_n)]|_{i\omega_n\rightarrow \omega+i0^+},\eqno{(2)}$$
where $\nu=0$, 1, $N$ is the number of sites in the lattice, and $G^{\nu\nu^\prime}_{{\bf k}{\bf k}^\prime}( i\omega_n)$ is the Green's function
in terms of the Matsubara frequencies and has the symmetry $G^{\nu\nu^\prime}_{{\bf k}{\bf k}^\prime}(\pm i\omega_n)\equiv G^{\nu^\prime\nu}_{{\bf k}^\prime{\bf k}}(\pm i\omega_n)$. To the first order in $V_s$, $\delta t$, $\delta \Delta_1$ and $\delta\Delta_2$, the Green's function
$G^{\nu\nu^\prime}_{{\bf k}{\bf k}^\prime}(i\omega_n)= G^0_{{\bf k}\nu}(i\omega_n)\delta_{{\bf k}{\bf k}^\prime}
\delta_{\nu\nu^\prime}+G^0_{{\bf k}\nu}(i\omega_n)G^0_{{\bf k}^\prime\nu^\prime}(i\omega_n)\{
\delta_{{\bf k}{\bf k}^\prime} {\cal D}_{\bf k}
[(-1)^{\nu^\prime} \xi_{{\bf k}\nu^\prime} f_{{\bf k}\nu}+ \xi_{{\bf k}\nu^\prime+1}g_{{\bf k}\nu}]\\ 
+\alpha_{{\bf k}{\bf k}^\prime}[(-1)^{\nu+\nu^\prime}\xi_{{\bf k}\nu}\xi_{{\bf k}^\prime\nu^\prime}-\xi_{{\bf k}\nu+1}\xi_{{\bf k}^\prime\nu^\prime+1}]+\beta_{{\bf k}{\bf k}^\prime}[(-1)^{\nu}\xi_{{\bf k}\nu}\\
\times \xi_{{\bf k}^\prime\nu^\prime+1}+(-1)^{\nu^\prime}\xi_{{\bf k}\nu+1}\xi_{{\bf k}^\prime\nu^\prime}]+{\cal D}_{{\bf k}^\prime}[(-1)^{\nu+\nu^\prime}\xi_{{\bf k}\nu}\xi_{{\bf k}^\prime\nu^\prime}\\
\times (\alpha_{{\bf k}{\bf k}^\prime}{\cal A}_{{\bf k}^\prime}+\beta_{{\bf k}{\bf k}^\prime}c_{{\bf k}^\prime})+(-1)^{\nu}\xi_{{\bf k}\nu}\xi_{{\bf k}^\prime\nu^\prime+1}(\alpha_{{\bf k}{\bf k}^\prime}b_{{\bf k}^\prime}+\beta_{{\bf k}{\bf k}^\prime}{\cal B}_{{\bf k}^\prime})+(-1)^{\nu^\prime}\xi_{{\bf k}\nu+1}\xi_{{\bf k}^\prime\nu^\prime}(-\alpha_{{\bf k}{\bf k}^\prime}c_{{\bf k}^\prime}+\beta_{{\bf k}{\bf k}^\prime}{\cal A}_{{\bf k}^\prime})+\xi_{{\bf k}\nu+1}\xi_{{\bf k}^\prime\nu^\prime+1}
(-\alpha_{{\bf k}{\bf k}^\prime}{\cal B}_{{\bf k}^\prime}+\beta_{{\bf k}{\bf k}^\prime}b_{{\bf k}^\prime})]+{\cal D}_{\bf k}[(u_{{\bf k}^\prime\nu^\prime}\alpha_{{\bf k}{\bf k}^\prime}+v_{{\bf k}^\prime\nu^\prime}\beta_{{\bf k}{\bf k}^\prime})(f_{{\bf k}\nu}X_{{\bf k}^\prime}+g_{{\bf k}\nu}Y_{{\bf k}^\prime})+(u_{{\bf k}^\prime\nu^\prime}\beta_{{\bf k}{\bf k}^\prime}-v_{{\bf k}^\prime\nu^\prime}\alpha_{{\bf k}{\bf k}^\prime})(f_{{\bf k}\nu}Y_{{\bf k}^\prime}+g_{{\bf k}\nu}Z_{{\bf k}^\prime})]\},$ 
$G^0_{{\bf k}\nu}(i\omega_n)=[i\omega_n-(-1)^{\nu}E_{\bf k}]^{-1}$,
$E_{\bf k}=\sqrt{({\epsilon}_{\bf k}-\mu)^2+\Delta ^2_{{\bf k}}}$,
$\xi^2_{{\bf k}\nu}=\frac{1}{2}[1+(-1)^\nu\frac{{\epsilon}_{\bf k}-\mu}{E_{\bf k}}]$,
$\xi_{{\bf k}\nu}\xi_{{\bf k}\nu+1}=\frac{\Delta_{\bf k}}{2E_{\bf k}}$,
$X_{\bf k}=\sum_{\nu}\xi^2_{{\bf k}\nu}G^0_{{\bf k}\nu}(i\omega_n)$,
$Y_{\bf k}=\sum_{\nu}(-1)^{\nu}\xi_{{\bf k}\nu}\xi_{{\bf k}\nu+1}G^0_{{\bf k}\nu}(i\omega_n)$,
$Z_{\bf k}=\sum_{\nu}\xi^2_{{\bf k}\nu+1}G^0_{{\bf k}\nu}(i\omega_n)$,
$\alpha_{{\bf k}{\bf k}^\prime}=\frac{V_s}{N}+\frac{2\delta t}{N}(\cos{k}_x+\cos{k}_y+\cos{k}_x^\prime+\cos{k}_y^\prime)$,
$\beta_{{\bf k}{\bf k}^\prime}=\frac{2\delta\Delta_1}{N}(\cos{k}_x-\cos{k}_y+\cos{k}_x^\prime-\cos{k}_y^\prime)+\frac{2\delta\Delta_2}{N}[\cos(k_x-2k^\prime_x)-\cos(k_x-k_x^\prime-k_y^\prime)-\cos(k_x-k_x^\prime+k_y^\prime)+\cos(k_y-k_x^\prime-k_y^\prime)+\cos(k_y-k_y^\prime+k_x^\prime)-\cos(k_y-2k_y^\prime)+ \cos(2k_x-k^\prime_x)
+\cos(k_x+k_y-k_y^\prime)-\cos(k_x+k_y-k_x^\prime)-\cos(2k_y-k_y^\prime)+\cos(k_x-k_y+k_y^\prime)
-\cos(k_x-k_y-k_x^\prime)]$,
$\lambda_{\bf k}=\frac{U+8V}{N}\sum_{{\bf k}^\prime}\xi^2_{{\bf k}^\prime 1}-\frac{2V}{N}
\sum_{{\bf k}^\prime}\xi^2_{{\bf k}^\prime 1}[\cos({k}_x^\prime-{k}_x)+\cos({k}_y^\prime-{k}_y)]$,
$\gamma_{\bf k}=-\frac{2V}{N}\sum_{{\bf k}^\prime}\xi_{{\bf k}^\prime 0}\xi_{{\bf k}^\prime 1}[\cos({k}_x^\prime-{k}_x)+\cos({k}_y^\prime-{k}_y)]$,
$a_{\bf k}=\lambda_{\bf k}X_{\bf k}+\gamma_{\bf k}Y_{\bf k}$,
$b_{\bf k}=-\lambda_{\bf k}Y_{\bf k}+\gamma_{\bf k}X_{\bf k}$,
$c_{\bf k}=\lambda_{\bf k}Y_{\bf k}+\gamma_{\bf k}Z_{\bf k}$,
$d_{\bf k}=-\lambda_{\bf k}Z_{\bf k}+\gamma_{\bf k}Y_{\bf k}$,
$f_{{\bf k}\nu}=(-1)^\nu\xi_{{\bf k}\nu}[\lambda_{\bf k}+(\lambda_{\bf k}^2+\gamma_{\bf k}^2)Z_{\bf k}]
+\xi_{{\bf k}\nu+1}[\gamma_{\bf k}-(\lambda_{\bf k}^2+\gamma_{\bf k}^2)Y_{\bf k}]$,
$g_{{\bf k}\nu}=(-1)^\nu\xi_{{\bf k}\nu}[\gamma_{\bf k}-(\lambda_{\bf k}^2+\gamma_{\bf k}^2)Y_{\bf k}]
+\xi_{{\bf k}\nu+1}[-\lambda_{\bf k}+(\lambda_{\bf k}^2+\gamma_{\bf k}^2)X_{\bf k}]$,
${\cal A}_{\bf k}=a_{\bf k}(1-d_{\bf k})+b_{\bf k}c_{\bf k}$,
${\cal B}_{\bf k}=d_{\bf k}(1-a_{\bf k})+b_{\bf k}c_{\bf k}$,
${\cal D}_{\bf k}=[(1-a_{\bf k})(1-d_{\bf k})-b_{\bf k}c_{\bf k}]^{-1}$,
$u_{{\bf k}\nu}=(-1)^\nu \xi_{{\bf k}\nu}(1+{\cal A}_{\bf k}{\cal D}_{\bf k})+\xi_{{\bf k}\nu+1}b_{\bf k}{\cal D}_{\bf k}$,
$v_{{\bf k}\nu}=(-1)^\nu \xi_{{\bf k}\nu}c_{\bf k}{\cal D}_{\bf k}+\xi_{{\bf k}\nu+1}(1+{\cal B}_{\bf k}{\cal D}_{\bf k})$.
Obviously, the weak magnetic potential $V_m$ doesnot appear in $G^{\nu\nu^\prime}_{{\bf k}{\bf k}^\prime}(i\omega_n)$ because it has only the higher order contributions to the Green's function. 

Here we point out that the Coulomb interaction itself doesnot produce the charge modulations in the present case. However, it has a strong impact on the LDOS induced by disorder.
Substituting the Green's function $G^{\nu\nu^\prime}_{{\bf k}{\bf k}^\prime}(\pm i\omega_n)$ into the formula (2), we can obtain the LDOS images at different energy and doping. In our calculations, we have chosen $N=400\times 400$, $\mu=-0.1246$ corresponding to about $15\%$ doping, $\Delta_0=44$ meV and $V_s=2\delta t=-2\delta\Delta_1=-4\delta\Delta_2$ for simplicity. Here we'd like to  mention that the other parameter options, including $\delta\Delta_1$ and $\delta\Delta_2$ to be positive and different doping, arrive at the similar conclusions. 

\begin{figure}
\rotatebox[origin=c]{0}{\includegraphics[angle=0, 
           height=1.4in]{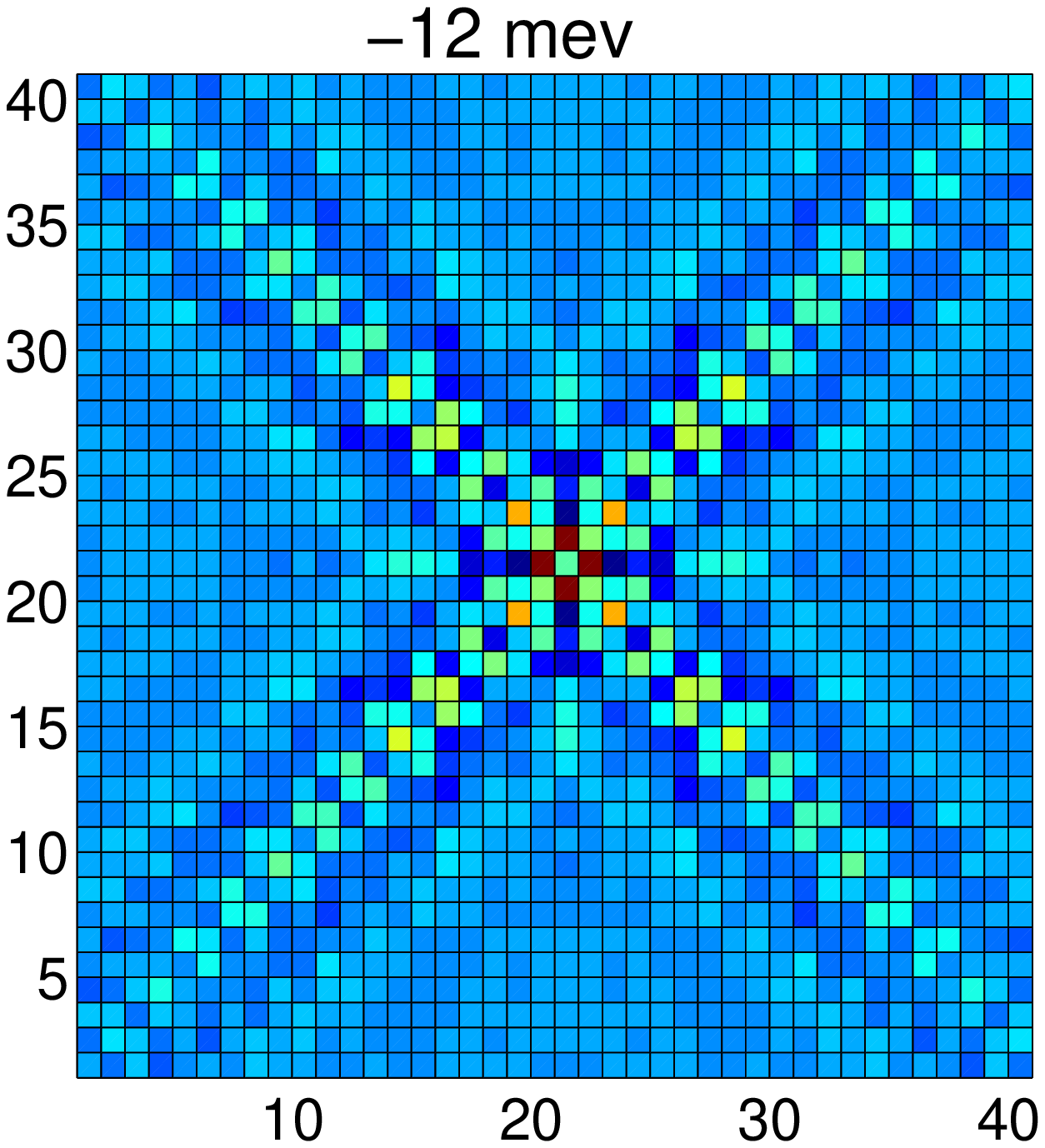}}
\rotatebox[origin=c]{0}{\includegraphics[angle=0, 
           height=1.4in]{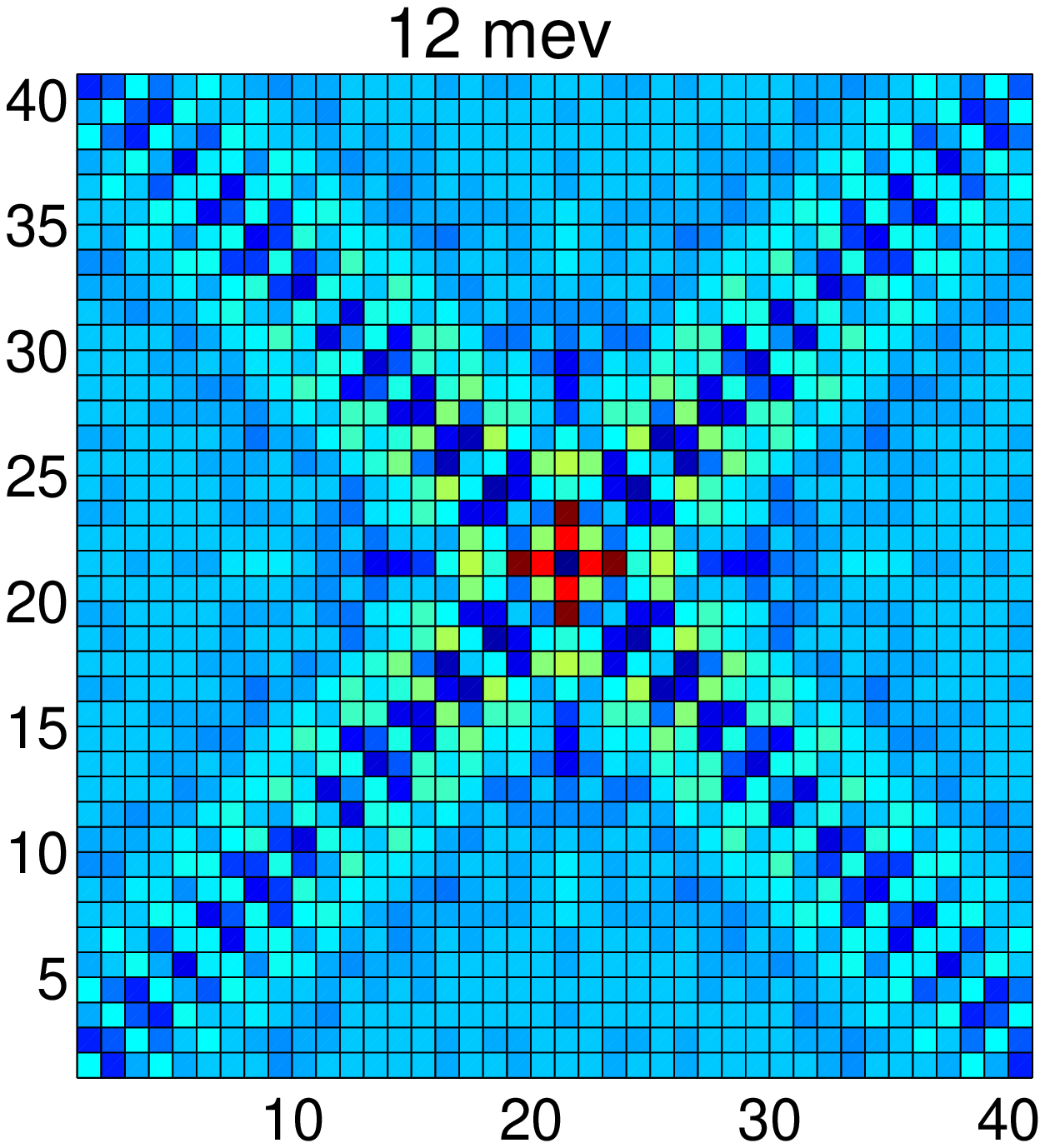}}
\rotatebox[origin=c]{0}{\includegraphics[angle=0, 
           height=1.4in]{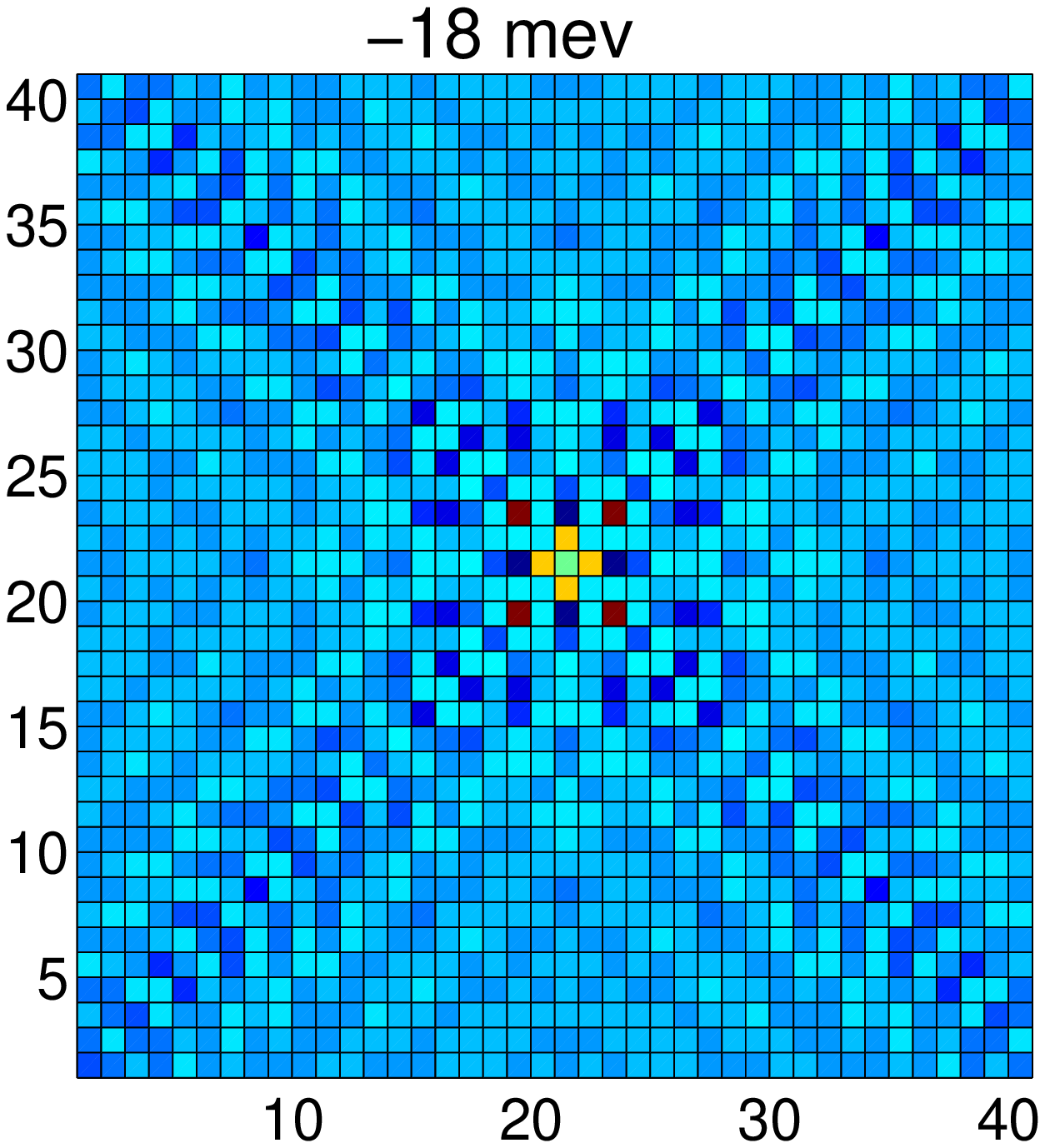}}
\rotatebox[origin=c]{0}{\includegraphics[angle=0, 
           height=1.4in]{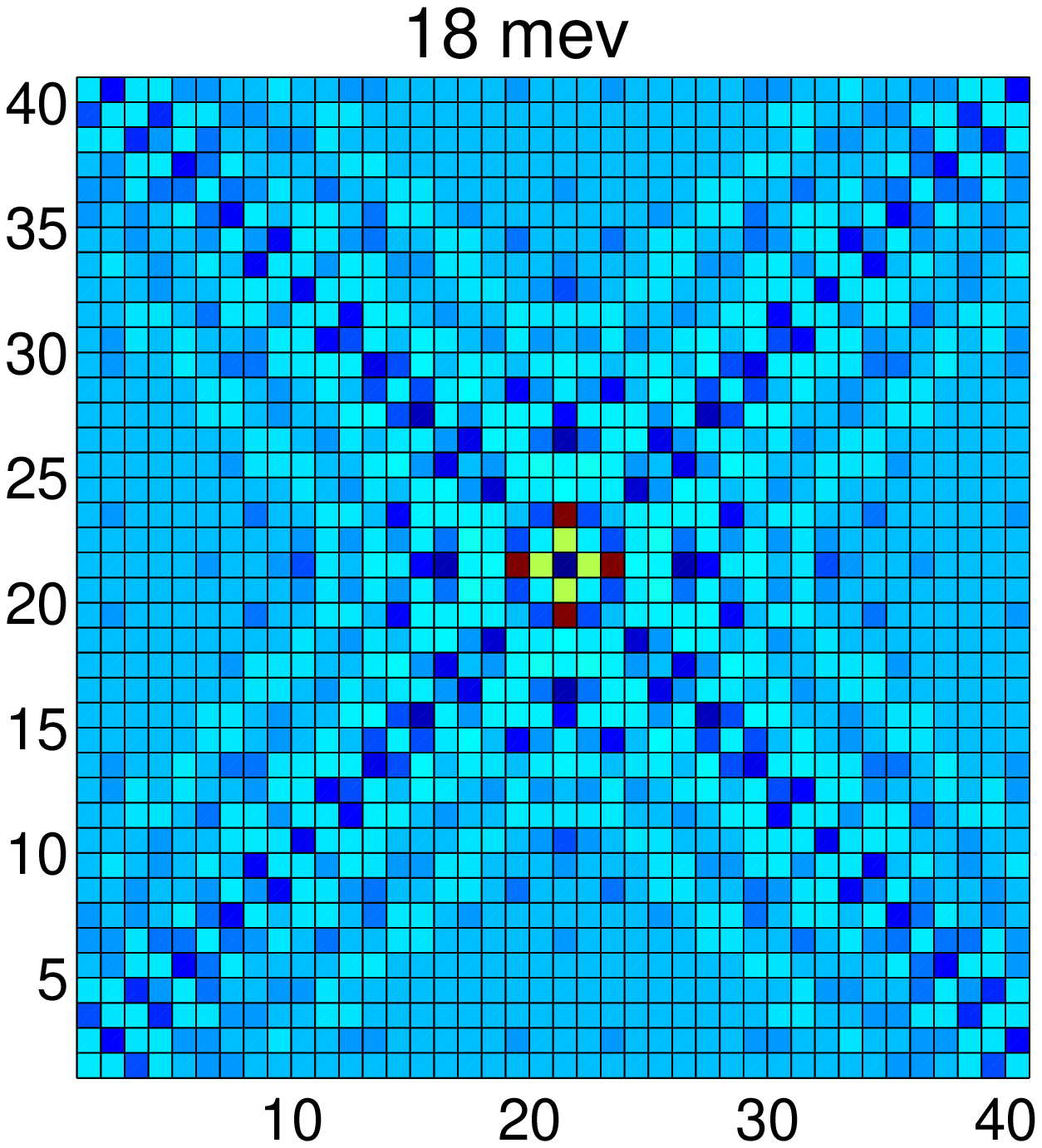}}
\rotatebox[origin=c]{0}{\includegraphics[angle=0, 
           height=1.4in]{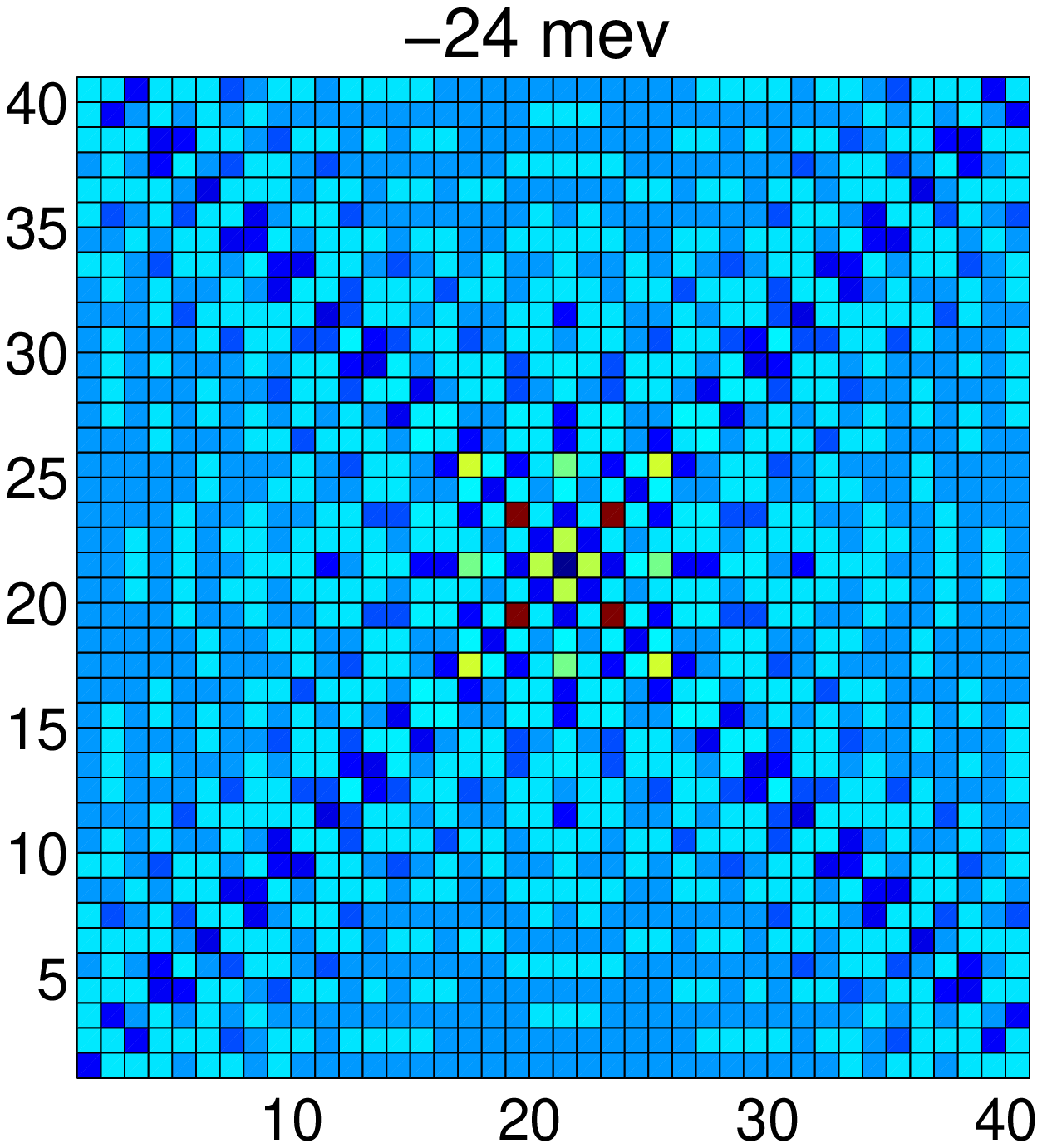}}
\rotatebox[origin=c]{0}{\includegraphics[angle=0, 
           height=1.4in]{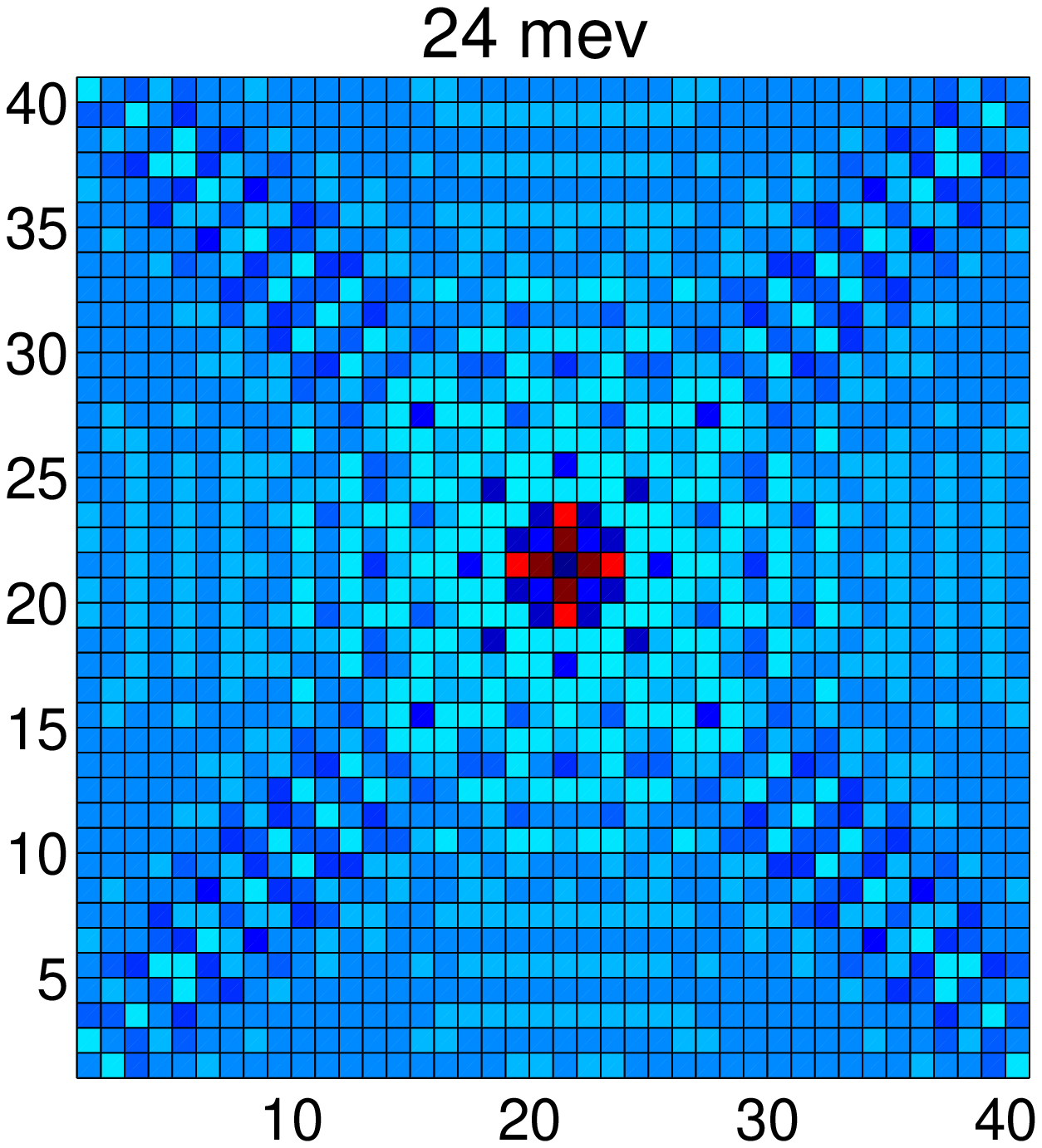}}  
\rotatebox[origin=c]{0}{\includegraphics[angle=0, 
           height=1.4in]{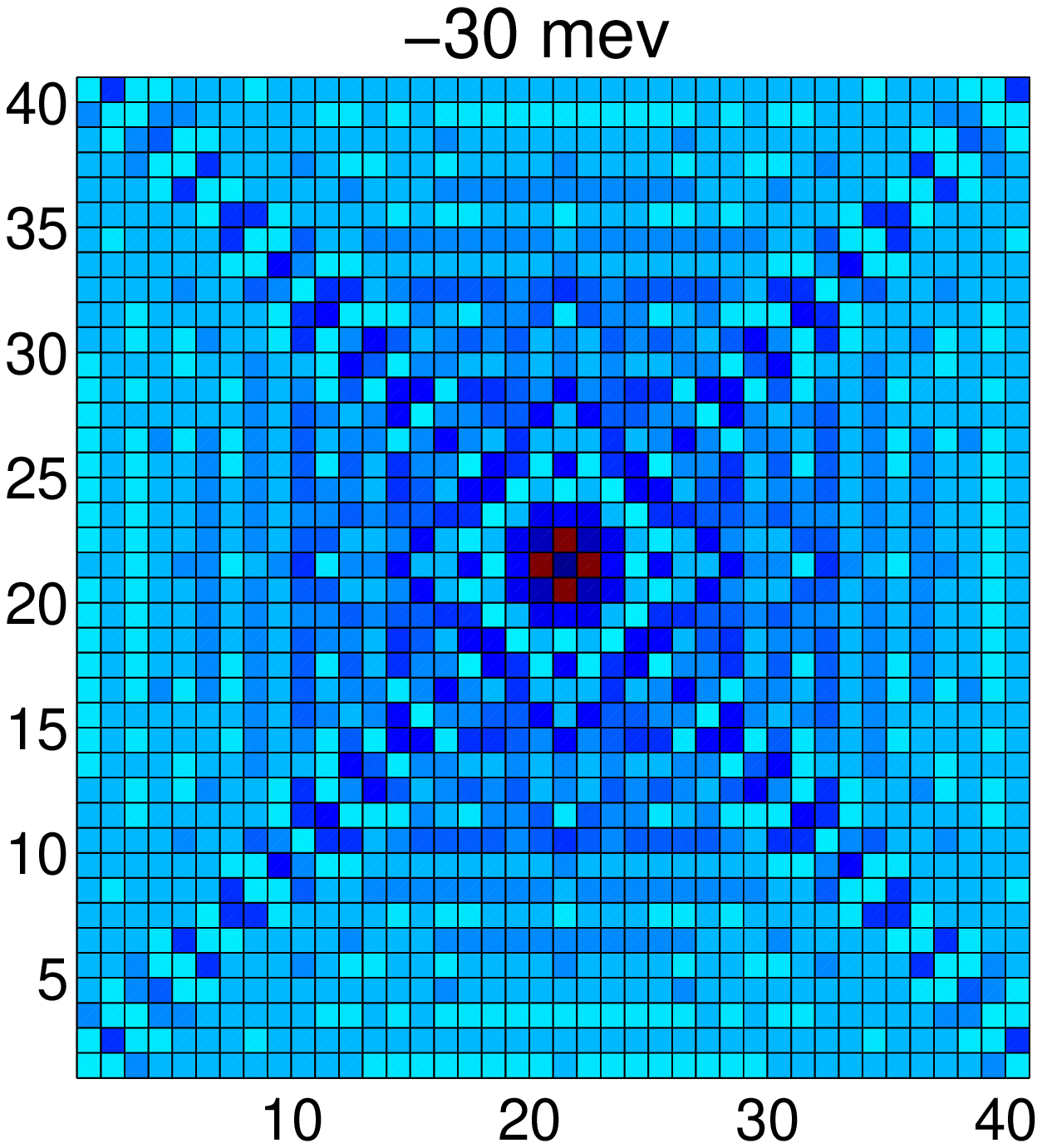}}
\rotatebox[origin=c]{0}{\includegraphics[angle=0, 
           height=1.4in]{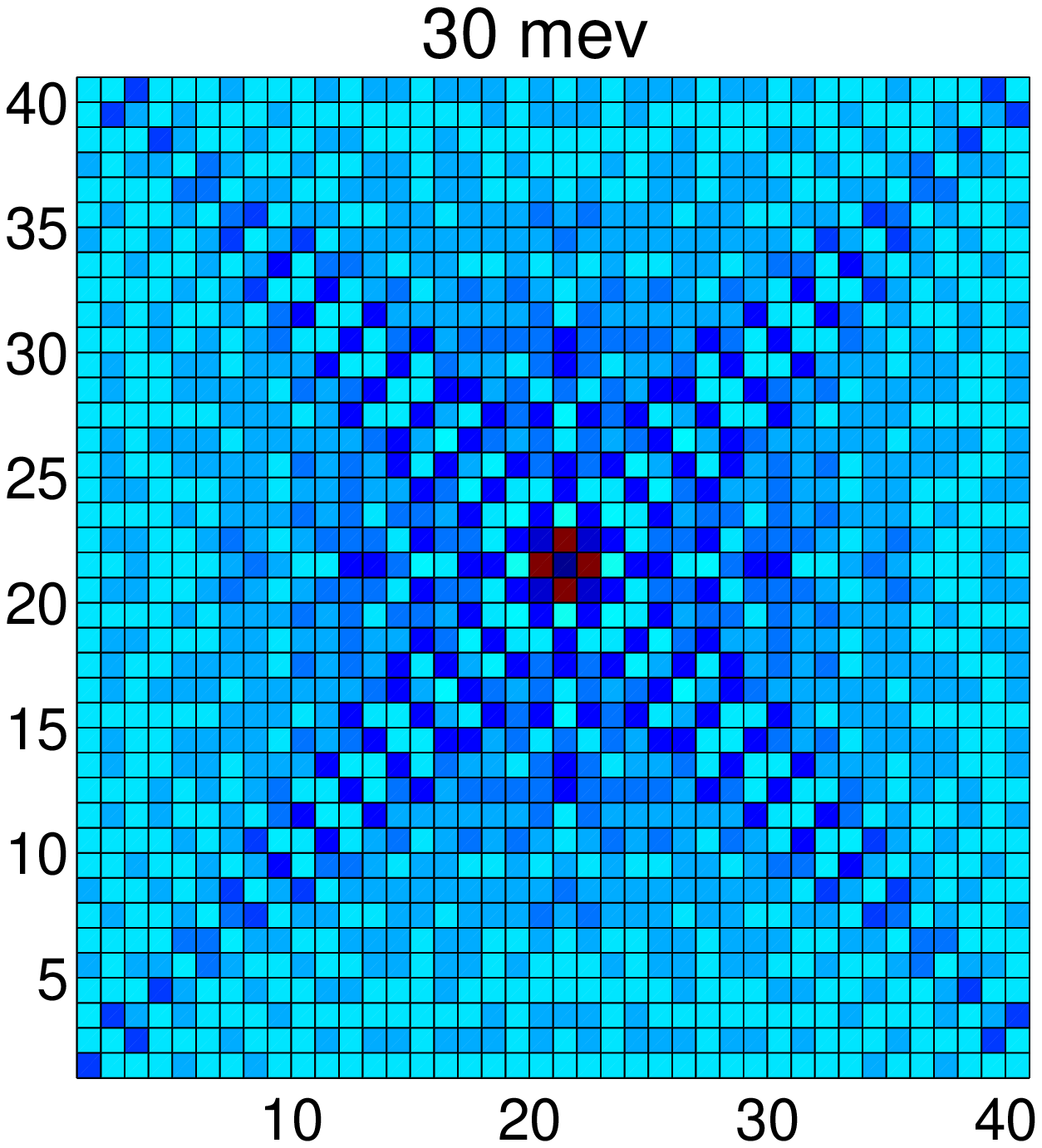}}                      
\caption {(Color online) The LDOS $\rho({\bf r},\omega)$ at different energy due to a single impurity at the center of a $40\times 40$ square. $U=0.25$ eV and $V=0.01$ eV.}
\end{figure}

Figure 1 shows the LDOS patterns with a moderate $U=0.25$ eV and a small $V=0.01$ eV at different energy due to a single impurity at the center of a $40\times 40$ square. In this case, the ratio of the on-site Coulomb interaction to the nearest neighbor hopping $\frac{U}{t}\approx 1.7$. Obviously, the LDOS patterns for $\omega>0$ have similarly periodic structures with those for $\omega<0$. At $\omega=\pm 12$ meV, the LDOS has a checkerboard modulation with a period $\sim 5a$. Its orientation is along $45^0$ to the Cu-O bonds. With increasing energy to $\pm 18$ meV, the modulation period becomes shorter to  $\sim 4a$. We note that there also exists an about $20\times 20$ checkerboard modulation along the Cu-O bonds around the impurity at $\pm 18$ meV. 
At high energy, e.g. $\omega=\pm 30$ meV, the LDOS shows a checkeroard pattern along the Cu-O bonds, which possesses a multi-period structure. Obviously, there doesnot exist the long-wavelength modulations. When $\omega=\pm 24$ meV, two kinds of the checkerboard patterns along both $45^0$ to the Cu-O bonds and the Cu-O bonds coexist. The charge modulation and its transformation with energy agree qualitatively the STM experiments. Obviously, the Coulomb interaction suppresses the maximum LDOS on or near the impurity sites and magnifies the Friedel oscillation in a large area at each energy [9].

\begin{figure}
\rotatebox[origin=c]{0}{\includegraphics[angle=0, 
           height=2.2in]{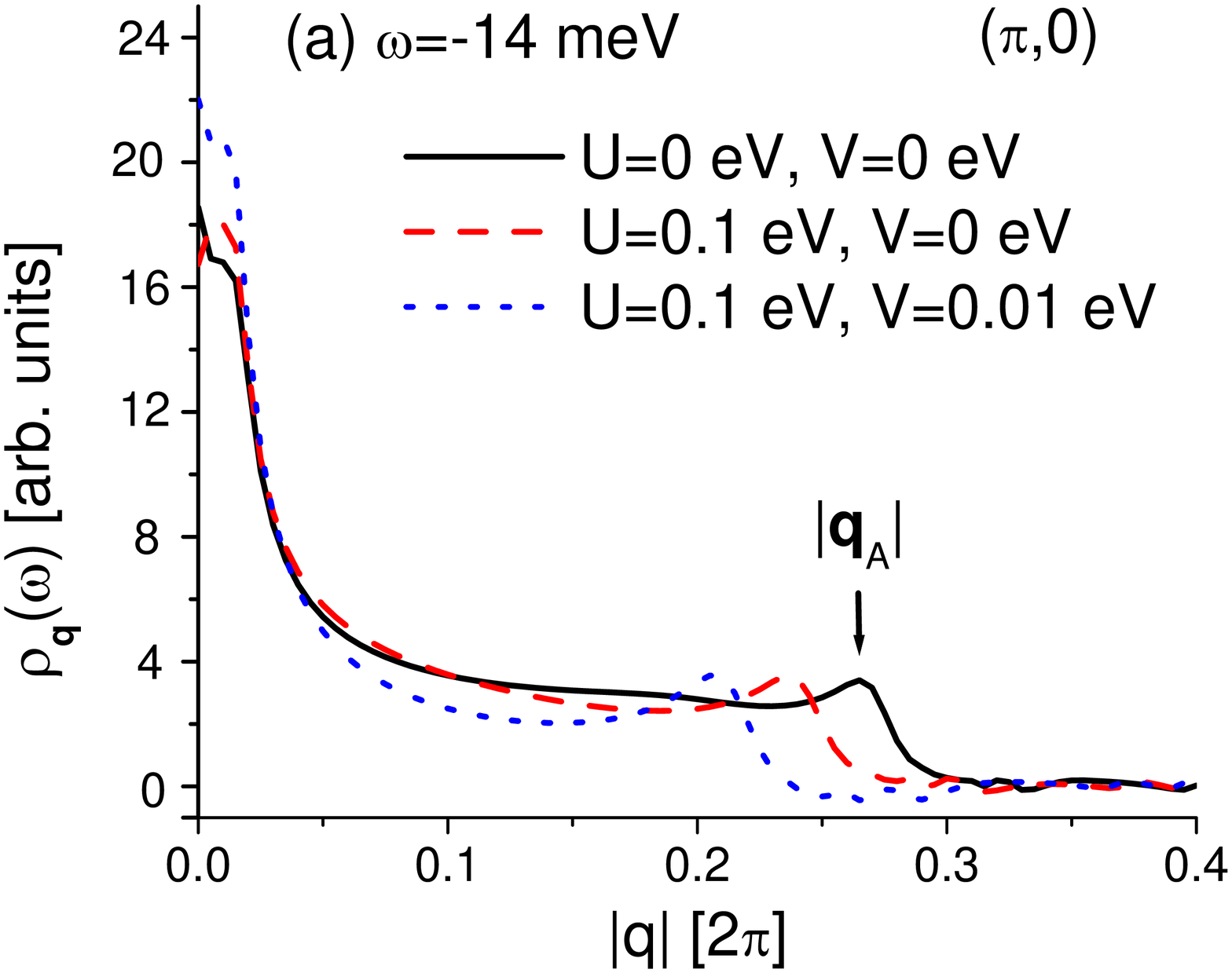}}
\rotatebox[origin=c]{0}{\includegraphics[angle=0, 
           height=2.2in]{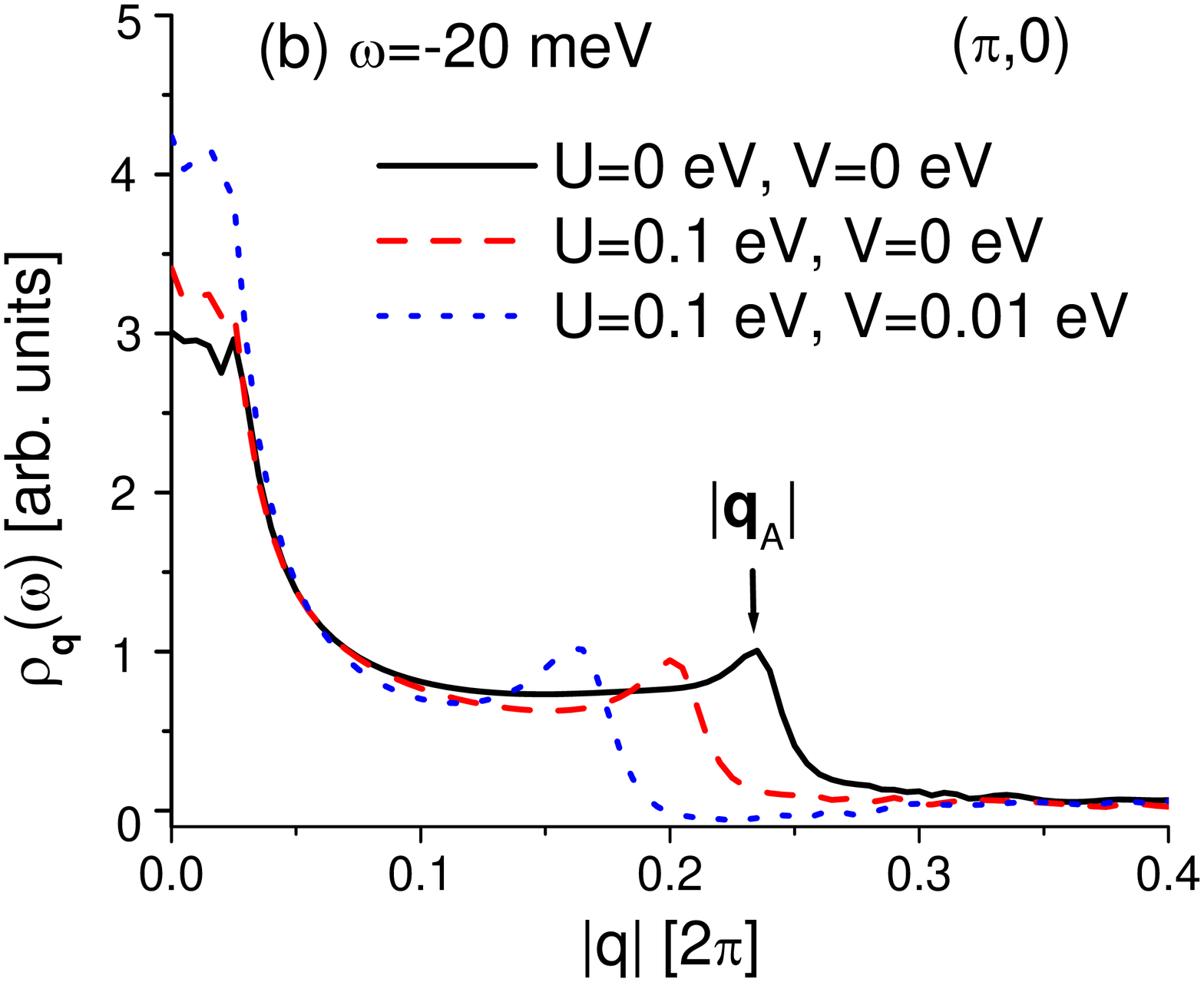}}
\caption{(Color online) The Fourier component $\rho_{\bf q}(\omega)$ of the LDOS along the antinodal direction at different energy and Coulomb interaction.}           
\end{figure}

In order to further understand the influence of the Coulomb interaction on the electronic states in the CuO$_2$ plane, we study the Fourier component of the LDOS, i.e. $\rho_{\bf q}(\omega)=\frac{1}{N}\sum_{\bf r}e^{-i{\bf q}\cdot{\bf r}}\rho({\bf r},\omega)$. In Fig. 2, we present $\rho_{\bf q}(\omega)$ along $(\pi,0)$ direction at different energy. Obviously, $\rho_{\bf q}(\omega)$ has a peak at the modulation wave vector $q_A$, which changes with $\omega$, $U$ and $V$. With increasing $U$ and $V$ ($\omega$ fixed) or $\omega$ ($U$ and $V$ fixed), the peak moves to the origin and $|q_A|$ becomes shorter. If $U$ or $V$ has an enough large value, the peak will disappear at high energy below the superconducting gap. This seems to be consistent with the STM experiments [1,2]. In contrast, the peak at $q_A$ produced by the pure quasiparticle interference never disappears when energy approaches to the superconducting gap.

\begin{figure}
\rotatebox[origin=c]{0}{\includegraphics[angle=0, 
           height=2.2in]{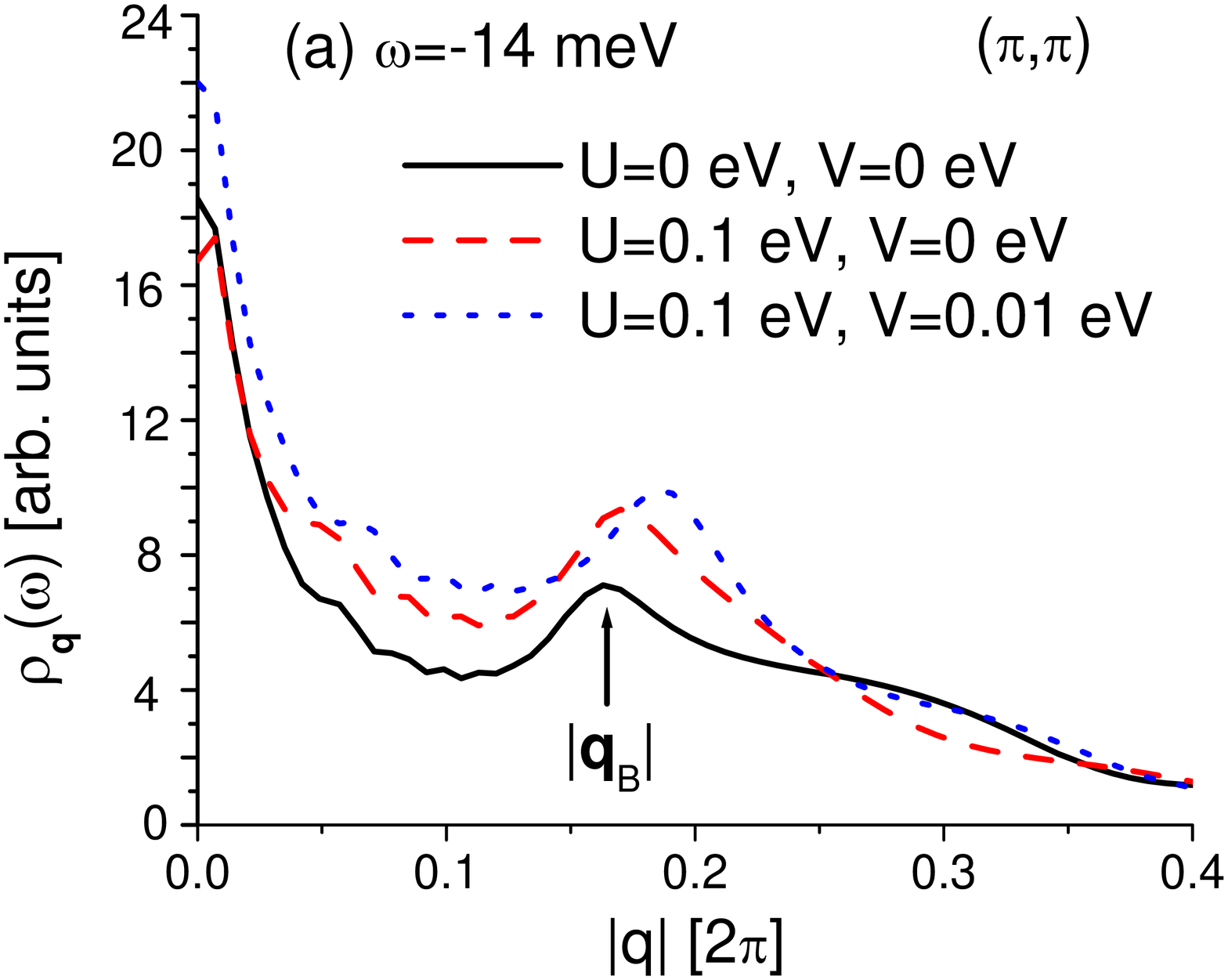}}
\rotatebox[origin=c]{0}{\includegraphics[angle=0, 
           height=2.2in]{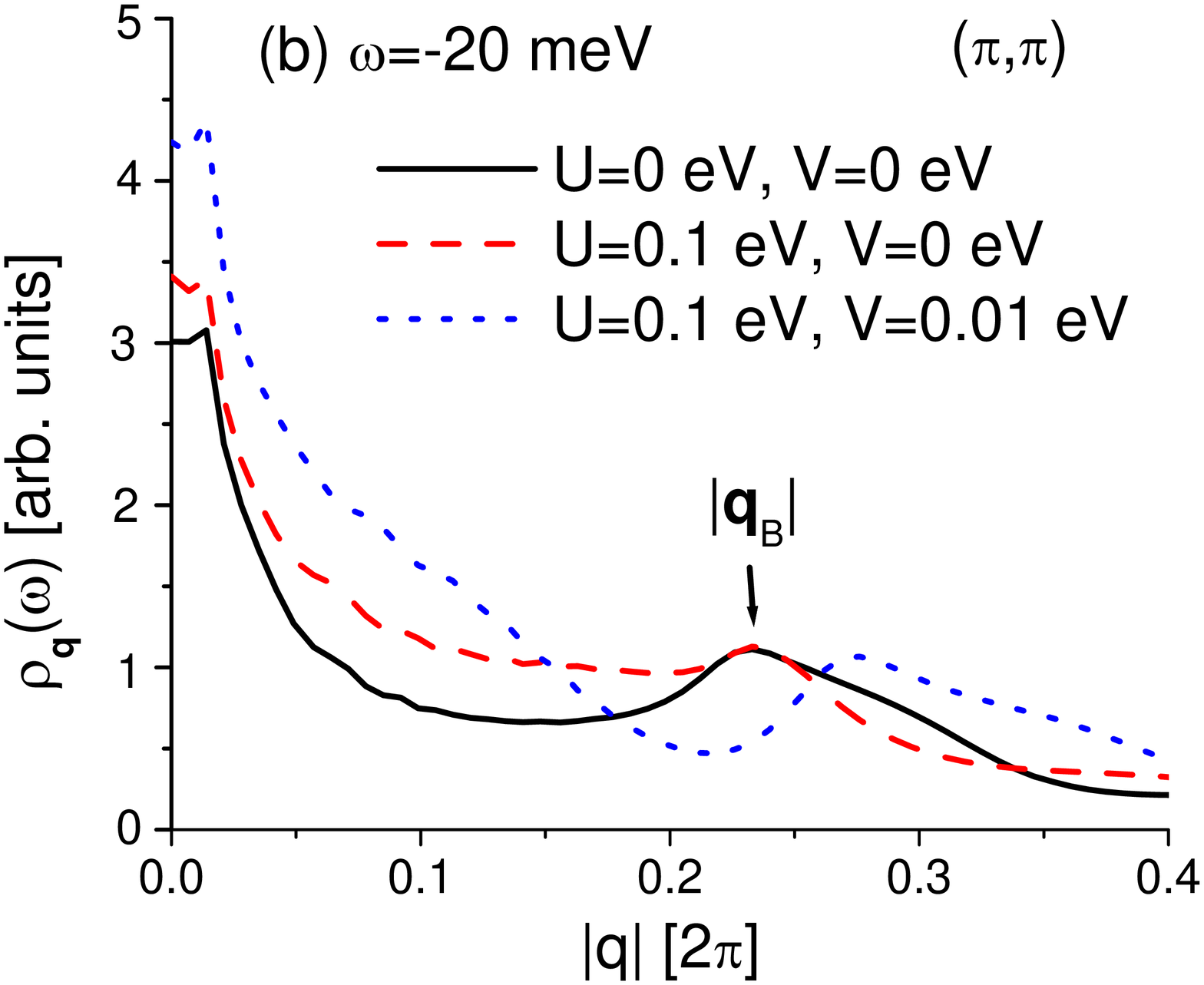}}
\caption{(Color online) The Fourier component $\rho_{\bf q}(\omega)$ of the LDOS  along the nodal direction at different energy and Coulomb interaction.}             
\end{figure}

Figure 3 shows $\rho_{\bf q}(\omega)$ along $(\pi,\pi)$ direction at different energy. We can see that when $U<\sim 0.1$ eV and $V=0$ eV, the peak at the modulation wave vector $q_B$ does not move with $U$ at each energy. However, $q_B$ becomes longer if $U>\sim 0.1$ eV or a small $V$ is presented. Therefore, when $U$ and $V$ are fixed, the peaks at $q_A$ and $q_B$ have an opposite shift with increasing energy. In other words, $q_A$ becomes shorter while $q_B$ becomes longer. The relation between $q_A (q_B)$ and $\omega$ has been observed by the STM experiments [1,2]. Note that both $q_A$ and $q_B$ are very sensitive to the presence of $V$ while a small $U$ term has a large impact on $q_A$.

In summary, we have investigated in detail the effect of the Coulomb interaction on sites and between the nearest neighbor sites on the electronic states in disordered cuprate superconductors. It is shown that the Coulomb interaction plays an important role in forming the checkerboard patterns of the LDOS modulations induced by the impurity. Our model (1) produces the essential features of the energy-dependent LDOS modulations as observed in the STM experiments [1,2,3]. In fact, cuprate superconductors contain a lot of random oxygen-dopant impurities. However, all the impurities correlate each other through the Coulomb interaction. We conclude that it is nothing but this kind of correlation that leads to the energy-dependent  checkerboard patterns of the LDOS modulations in cuprate superconductors.

The author would like to thank Professor C. S. Ting and Professor S. H. Pan for useful discussions.
This work was supported by the Texas Center for Superconductivity at the University of Houston 
and by the Robert A. Welch Foundation under grant no E-1146.


\end{document}